\documentstyle[aps]{revtex}

\begin{document}
\title{The homogeneous balance method, Lax pair, Hirota transformation and a
general fifth-order KdV equation}
\author{Yang Lei, Zhang Fajiang, Wang Yinghai}
\address{{\it Department of physics, Lanzhou University, Gansu 730000, China}}
\maketitle

\begin{abstract}
In this paper, some notes of the homogeneous balance (HB) method are
discussed by a kind of general fifth-order KdV (fKdV) equation. Frist, the
auto-B\"{a}cklund transformation and lax represents of the higher-order KdV
equation(a specific forms of fKdV equation) are obtained by the HB method.
Then, the connection of the Hirota transformation and the HB method is
discussed by two specific fKdV equations: the Sawada-Kotera equation and the
Lax's equation. At the same time, the solitary wave solution of the general
fifth-order KdV equation is obtained.
\end{abstract}

From 70's, a vast variety of the simple and direct methods to find analytic
solutions of PDE have been developed. Recently, the homogeneous balance (HB)
method has drawn lots of interests in seeking the solitary wave solution[1]
and other kinds of solutions[2], the solitary wave solution of some
nonlinear PDEs[3] have been obtained by this method and the relation between
the HB method and the CK method or B\"{a}cklund transformation has been
discussed[4]. Then how doese the HB method connect with Lax pair and Hirota
transformation? In this letter, the relation between the HB method and other
methods is discussed by a general fifth-order KdV equation. At the same
time, the solitary wave solution of a general KdV equation is obtained.

The general form of the fifth-order KdV is written as

\begin{equation}
u_t+au_{xxxxx}+buu_{xxx}+cu_xu_{xx}+du^2u_x=0,  \label{1}
\end{equation}
where $a$, $b$, $c$ and $d$ are constants. The equation (1) includes lots of
equations which have been studied widely [5], when the special values of $a$%
, $b$, $c$, $d$ are set. According to  the HB method, first, supposing the
transformation of $u$, 
\begin{equation}
u=\sum_{i=0}^Nf^{(i)}(\omega (x,t)),  \label{2}
\end{equation}
then substituting (2) into (1), and balancing the nonlinear term $%
buu_{xxx}+cu_xu_{xx}+du^2u_x$ and the linear term $au_{xxxxx}$, $N$ can be
decided to be $2$, thus the transformation (2) is order to be

\begin{equation}
u=f^{^{\prime \prime }}\omega _x^2+f^{^{\prime }}\omega _{xx}.  \label{3}
\end{equation}
Substituting (3) into (1), and collecting all homogeneous terms in partial
dericatives of $\omega (x,t)$, we have (because the formula is so long, just
one part of it is shown here)

\begin{equation}
(d(f^{\prime \prime })^2f^{(3)}+cf^{(3)}f^{(4)}+bf^{\prime \prime
}f^{(5)}+af^{(7)})\omega _x^7+\ldots \ldots =0.  \label{4}
\end{equation}
Setting the coefficient of $\omega _x^7$ in (4) to zero yields an ordinary
differential equation for $f$

\begin{equation}
d(f^{\prime \prime })^2f^{(3)}+cf^{(3)}f^{(4)}+bf^{\prime \prime
}f^{(5)}+af^{(7)}=0.  \label{5}
\end{equation}
Solving (5), the solution of the equation(5) is

\begin{equation}
f=\alpha \ln \omega ,  \label{6}
\end{equation}
where $\alpha =\frac{6b+3c\pm 3\sqrt{(2b+c)^2-40ad}}d$, it's obviously that
the parameter $a$, $b$, $c$, $d$ need to satisfy $(2b+c)^2-40ad>0$. The
solution(6) yields

\begin{equation}
f^{(3)}f^{(4)}=-\frac \alpha {60}f^{(7)},\ \ f^{\prime \prime }f^{(5)}=-%
\frac \alpha {30}f^{(7)},\ \ (f^{\prime \prime })^2f^{(3)}=\frac \alpha {360}%
f^{(7)}\ldots \ldots ,  \label{7}
\end{equation}
Substituting (7) into (4), formula (4) can be simplified to a linear
polynomial of $f^{\prime },f^{\prime \prime },\ldots $, then setting the
coefficients of $f^{\prime },f^{\prime \prime },\ldots $ to zero yields a
set of partial differential equations for $\omega (x,t)$,

\begin{equation}
\begin{array}{l}
\omega _{xxt}+a\omega _{7x}=0, \\ 
-2\omega _x\omega _{xt}-\omega _t\omega _{xx}+(c\alpha -35a)\omega
_{xxx}\omega _{4x}+(b\alpha -21a)\omega _{xx}\omega _{5x}-7a\omega _x\omega
_{6x}=0, \\ 
2\omega _t\omega _x^2+(210a-10b\alpha -3c\alpha +d\alpha ^2)\omega
_{xx}^2\omega _{xxx}+(140a-4c\alpha )\omega _x\omega _{xxx}^2+(210a-5b\alpha
-3c\alpha )\omega _x\omega _{xx}\omega _{4x} \\ 
+(42a-b\alpha )\omega _x^2\omega _{5x}=0, \\ 
(630a-30b\alpha -9c\alpha +3d\alpha ^2)\omega _{xx}^3+(1260a-30b\alpha
-24c\alpha +2d\alpha ^2)\omega _x\omega _{xx}\omega _{xxx}+(210a-5b\alpha
-2c\alpha )\omega _x^2\omega _{4x}=0, \\ 
(2520a-90b\alpha -42c\alpha +8d\alpha ^2)\omega _{xx}^2+(840a-20b\alpha
-14c\alpha +d\alpha ^2)\omega _x\omega _{xxx}=0,
\end{array}
\label{8}
\end{equation}
Where the equation $\omega _{xxt}+a\omega _{7x}=0$ which is from the
coefficients of $f^{\prime }$ is a linear PDE. It's easy to know this
equation has the travelling wave solution

\begin{equation}
\omega (x,t)=c_0+c_1e^{\beta (x-vt)},  \label{9}
\end{equation}
where $c_0$, $c_1$and $v$ are arbitrary constants, $\beta =-(\frac va)^{%
\frac 14}$. Substituting formula(9) into the set of equations (8), a set of
nonlinear algebraic equations are gotten. Solving the set of nonlinear
algebraic equations, the parameter $a$, $b$, $c$, $d$ need to satisfy the
formula

\begin{equation}
-4b^3-8b^2c-5bc^2-c^3+40abd+30acd+(10ad-c^2-3bc-2b^2)\sqrt{(2b+c)^2-40ad}=0.
\label{10}
\end{equation}
Then the exact solitary wave solution of the general fifth KdV equation is
obtained,

\begin{equation}
u(x,t)=\frac{3c_0c_1(2b+c\pm \sqrt{(2b+c)^2-40ad})\sqrt{v}e^{(\frac va)^{%
\frac 14}(x-vt)}}{\sqrt{a}d(c_1+c_0e^{(\frac va)^{\frac 14}(x-vt)})^2}.
\label{11}
\end{equation}

In this section, by the HB method, the auto-B\"{a}cklund transformation and
lax represents of a special fKdV equation[5] are obtained. The fKdV equation
is

\begin{equation}
u_t-\frac 14u_{xxxxx}-\frac 52uu_{xxx}-5u_xu_{xx}-\frac{15}2u^2u_x=0,
\label{12}
\end{equation}
where $a=-\frac 14$, $b=-\frac 52$, $c=-5$, $d=-\frac{15}2$. Base on the
transformation (3), ordering

\begin{equation}
u=f^{^{\prime \prime }}\omega _x^2+f^{^{\prime }}\omega _{xx}+v,  \label{13}
\end{equation}
where $v$ is a function of $x$, $t$, repeating the step (4)-(7), we yield a
set of partial differential equations (14-18) for $\omega (x,t)$ and $v(x,t)$

\begin{equation}
-60vv_x\omega _{xx}+4\omega _{xxt}-10\omega _{xx}v_{xxx}-30v^2\omega
_{xxx}-20v_{xx}\omega _{xxx}-20v_x\omega _{4x}-10v\omega _{5x}-\omega
_{7x}=0,  \label{14}
\end{equation}

\begin{equation}
\begin{array}{l}
60vv_x\omega _x^2-8\omega _x\omega _{xt}-4\omega _t\omega _{xx}+90v^2\omega
_x\omega _{xx}+60\omega _xv_{xx}\omega _{xx}+10\omega
_x^2v_{xxx}+80v_x\omega _x\omega _{xxx} \\ 
-20v\omega _{xx}\omega _{xxx}+50v\omega _x\omega _{4x}-5\omega _{xxx}\omega
_{4x}+\omega _{xx}\omega _{5x}+7\omega _x\omega _{6x}=0,
\end{array}
\label{15}
\end{equation}

\begin{equation}
\begin{array}{l}
4\omega _t\omega _x^2-30v^2\omega _x^3-20\omega _x^3v_{xx}-60v_x\omega
_x^2\omega _{xx}+30v\omega _x\omega _{xx}^2-40v\omega _x^2\omega
_{xxx}-5\omega _{xx}^2\omega _{xxx}+10\omega _x\omega _{xxx}^2 \\ 
+5\omega _x\omega _{xx}\omega _{4x}-11\omega _x^2\omega _{5x}=0,
\end{array}
\label{16}
\end{equation}

\begin{equation}
2v_x\omega _x^3+\omega _{xx}^3-2\omega _x\omega _{xx}\omega _{xxx}+\omega
_x^2\omega _{4x}=0,  \label{17}
\end{equation}
and $v(x,t)$ satisfies the equation

\begin{equation}
v_t-\frac 14v_{xxxxx}-\frac 52vv_{xxx}-5v_xv_{xx}-\frac{15}2v^2v_x=0,
\label{18}
\end{equation}
which is the same as the equation (12) of $u(x,t)$. Thus, the set of
equation (13-18) constitute an invariant-B\"{a}cklund transformation of the
fKdV equation (12). Considering the equation (17), the $v_x(x,t)$ can be
solved

\begin{equation}
v_x(x,t)=-\frac{\omega _{xx}^3-2\omega _x\omega _{xx}\omega _{xxx}+\omega
_x^2\omega _{xxxx}}{\omega _x^3},  \label{19}
\end{equation}
then integrating the formula (19), we have

\begin{equation}
v(x,t)=\frac 12\frac{\omega _{xx}^2}{\omega _x^2}-\frac{\omega _{xxx}}{%
\omega _x}+\lambda   \label{20}
\end{equation}
where $\lambda $ is a integration constant, here $\frac 12\frac{\omega
_{xx}^2}{\omega _x^2}-\frac{\omega _{xxx}}{\omega _x}$ just is the
Schwarzian derivative $S_x\omega (x,t)$, namely $v(x,t)=S_x\omega
(x,t)+\lambda $. Substituting the formula(20) and the transformation(13)
into the equation(12), the equation(12) can be written to be

\begin{equation}
4\omega _t-\omega _x(S_x\omega (x,t))_{xx}-\frac 32\omega _x(S_x\omega
(x,t))^2-10\lambda \omega _xS_x\omega (x,t)-30\lambda ^2=0.  \label{21}
\end{equation}
According the equation(21), which is the form of the Schwarzian derivative
of the equation (12), then it's readily to obtain the Lax pair[5]

\[
\nu _{xx}=\alpha \nu ,\qquad \nu _t=\beta \nu _x+\gamma \nu , 
\]
where

\[
\alpha =\lambda -u,\qquad \beta =\frac 12u_{xx}+\frac 32u^2+2\lambda
u+4\lambda ^2,\qquad \gamma =-\frac 12\beta _x+\alpha , 
\]
Applied the HB method, the Schwarzian derivative forms of some nonlinear
PDEs can be obtained, and the lax representation of the PDEs can be gotten,
such as the KdV, mKdV and the higher-order KdV.

In this section, the connection between Hirota transformation and HB method
is discussed. In 1970's, Hirota developed a method to obtain the special
solution of nonlinear PDE, namely, Hirota transformation which can transform
a nonlinear PDE to a bilinear form. For example, the Hirota transformation $%
u=2(\lg f)_{xx}$ can transfer the KdV equation to $[D_x(D_t+D_x^3)+\lambda
]f\cdot f=0$, where $D$ is Hirota's binary operator. It's easy to verify
that the Hirota transformation $u=2(\lg f)_{xx}$ can be obtained from the HB
method. Then can the Hirota transformation of the PDEs be obtained from the
HB method? In this paper, two special forms of the equation (1) (the
Sawada-Kotera equation and the Lax's equation) are discussed. The
Sawada-Kotera(SK) equation[5] is

\begin{equation}
u_t+u_{xxxxx}+15uu_{xxx}+15u_xu_{xx}+45u^2u_x=0.  \label{22}
\end{equation}
Substituting $a=1$, $b=15$, $c=15$, $d=45$ into the formula (6), the
transformation (3) can be written to $u=2(\lg f)_{xx}$, which is the Hirota
transformation for the SK equation. By this transformation the SK equation
is transformed into bilinear form $[D_x(D_t+D_x^5)]f\cdot f=0$. The Lax's
equation[5] is

\begin{equation}
u_t+u_{xxxxx}+10uu_{xxx}+20u_xu_{xx}+90u^2u_x=0.  \label{23}
\end{equation}
By the Hirota transformation $u=2(\lg \omega )_{xx}$, the Lax's equation is
transformed into bilinear form $[D_x(D_t+D_x^5)-\frac 53D_t(D_t+D_x^3)]f%
\cdot f=0$. Substituting $a=1$, $b=10$, $c=20$, $d=90$ into the formula (6),
the constrain condition $(2b+c)^2-40ad>0$ can't be satisfied. This means the
equation (5) has no real solution, and the real transformation (3) is
absent. So the HB method is invalid to the Lax's equation, and the Hirota
transformation $u=2(\lg \omega )_{xx}$ can't be obtained by HB method. These
results show that the HB method can't be use to obtain the Hirota
transformation, even in a kind of nonlinear PDE, but for some famous
equations(KdV, mKdV, KP, Bugers and {\it etc}) the Hirota transformation can
be obtained by the HB method.

In summary, a kind of general fifth-order KdV equation, which includes lots
of fifth-order KdV [5], is investigated in this paper. The solitary wave
solution of the fKdV equation is obtained. The auto-B\"{a}cklund
transformation and lax represents(which has been obtain by other methods) of
a integrate PDE(the higher-order KdV equation) are obtained by the HB
method. And the relation of the Hirota transformation and HB method is
discussed.

{\bf Acknowledgments: }This work was supported by the National Natural
Science Foundation of China. The corresponding author email address:
yhwang@lzu.edu.cn (Wang Yinghai).

\end{document}